\documentclass[10pt]{article}
\usepackage[utf8]{inputenc}
\usepackage{graphicx}
\usepackage[nohead,margin=1.0in]{geometry}

\usepackage{amsmath}
\usepackage{amsthm}
\usepackage{amsfonts}

\usepackage{booktabs}
\usepackage{graphicx}
\usepackage{xcolor}
\usepackage{pgfplots}

\pgfplotsset{compat=1.8}
\usetikzlibrary{positioning,calc,arrows,arrows.meta,shapes,shapes.multipart,pgfplots.statistics}

\usepgfplotslibrary{statistics}
\pgfplotsset{compat=1.6}
\pgfplotsset{
	matrixaxis/.style={
    	scale only axis,
        view={0}{90},
        enlargelimits=false,
        axis on top,
        axis equal image,
		colormap/blackwhite,
        point meta max=0.5,
        point meta min=0.0,
        colorbar style={
        	yticklabel style={
                /pgf/number format/.cd,
                fixed,
                precision=2,
                fixed zerofill,
            },
            ylabel={NRMSE},
            ylabel style={
            	yshift=5mm,
                rotate=180,
            },
        },
    },
}

\newcommand{\N}{\mathbb{N}}

\newcommand{\R}{\mathbb{R}}
\newcommand{\C}{\mathbb{C}}
\newcommand{\myEqref}[1]{(\ref{#1})}

\def\freqa{5M5}
\def\freqaStr{$5,5$}
\def\freqb{8M8}
\def\freqbStr{$8,8$}

\begin{document}

\title{Towards Accurate Modeling of the Multidimensional Magnetic Particle Imaging Physics}

\author{T. Kluth$^1$, P. Szwargulski$^{2,3}$, T. Knopp$^{2,3}$}
\date{{\footnotesize $^{1}$Center for Industrial Mathematics, University of Bremen\\$^{2}$Section for Biomedical Imaging, University Medical Center Hamburg-Eppendorf\\$^{3}$Institute for Biomedical Imaging, Hamburg University of Technology, Germany}}
\maketitle
\begin{abstract}
The image reconstruction problem of the tomographic imaging technique magnetic particle imaging (MPI) requires the solution of a linear inverse problem. One prerequisite for this task is that the imaging operator that describes the mapping between the tomographic image and the measured signal is accurately known. For 2D and 3D excitation patterns, it is common to measure the system matrix in a calibration procedure, that is both, very time consuming and adds noise to the operator. The need for measuring the system matrix is due to the lack of an accurate physical model that is capable of describing the nanoparticles' magnetization behavior. Within this work we introduce a physical model that is based on N\'{e}el rotation for large particle ensembles and we find model parameters that describe measured 2D MPI data with much higher precision than state of the art MPI models. With phantom experiments we show that the simulated system matrix can be used for image reconstruction and reduces artifacts due to model-mismatch considerably.
\end{abstract}

\subsubsection*{A. Introduction} %\paragraph{Introduction}
Tomographic imaging techniques play a major role in medical diagnostics and treatment. In 2005, a new method called magnetic particle imaging (MPI) was introduced \cite{Gleich2005}, which allows to image magnetic nanoparticles (MNPs) with high sensitivity and high spatio-temporal resolution. In contrast to MRI, the contrast in MPI is positive and the reconstructed images are quantitative. %\cite{zheng2015magnetic}. 
With these properties MPI is an interesting candidate for various medical applications ranging from vascular imaging \cite{vaalma2017magnetic} to targeted imaging  \cite{yu2017magnetic}. %\cite{Weizenecker2007,vogel2016first,graeser2017towards,vaalma2017magnetic} including interventional procedures \cite{haegele2016magnetic,salamon2016magnetic,herz2018magnetic} over to targeted imaging \cite{bulte2015quantitative,yu2017magnetic}. In particular in the field of neuroimaging recently various works have been published \cite{mason2017design,cooley2018rodent,wu2019review,ludewig2017magnetic}.

%For a detailed overview on the basic principle of MPI we refer the reader to \cite{knopp2017magnetic}. Within this work, we recapture that the 

Signal encoding in MPI is achieved by exciting the MNP with one or several homogeneous excitation fields with a frequency in the kHz region \cite{knopp2017magnetic}. The MNP respond with a change of their magnetization, which is measured with one or multiple receive coils. Spatial encoding is achieved by superimposing a static gradient field leading to a spatially dependent magnetization response of the MNP to the excitation field. In ferrofluids the change of the particle magnetization is enabled by two different dynamic processes \cite{Kluth2018a}: The Brownian rotation describes the mechanical alignment of the entire particle with a change of the magnetic field whereas N\'{e}el rotation describes the alignment of the particle's inner magnetic moment. Note that due to the continuous excitation the system does not relax in MPI. 
These dynamics in the context of MPI were initially studied by Weizenecker et al. \cite{Weizenecker2010particle} in terms of stochastic ordinary differential equations. A first attempt to Brownian rotation via the Fokker-Planck equation was presented by Yoshida and Enpuku for 1D excitation patterns \cite{Yoshida2012} followed by further works in this direction \cite{weizenecker2012micro,Yoshida2012b,yoshida2013characterization,rogge2013simulation,reeves2014approaches}. 
Different MNP are often categorized by the dominating mechanism. For example, Resovist's magnetization behavior is mainly determined by N\'{e}el rotation, whereas a frequency-dependent (with respect to the applied field) influence of Brownian rotation can be observed \cite{ludwig2013characterization}. 
Furthermore, the orientation of the particles' easy axis significantly influences the magnetization behavior due to the N\'{e}el rotation \cite{Yoshida2017}.
The still incomplete nature of models for imaging in MPI has motivated an increasing number of studies in the context of Brownian and N\'{e}el rotations for MPI excitation patterns \cite{weizenecker2012micro,Martens2013,Deissler2014,Enpuku2014,Shah2015,graeser2015trajectory}.

%The Brownian rotation model was experimentally validated using low frequencies in one-dimensional sinusoidial excitation patterns \cite{Martens2013}.
%&Furthermore, magnetization dynamics can substantially differ such that a distinction between different kinds of tracers is possible \cite{rahmer2015first}. 
%The use of these particle models is of particular interest, but they have not been applied extensively to the imaging problem. 

%After measuring the voltage induced in the receive coils, the raw data needs to be transformed into image domain. To this end, one needs to invert the physical process that caused the measured signal. Based on Farady's law of induction, it can be shown that the MPI imaging operator is linear and can be written as a Fredholm integral equation of the first kind. The integral kernel is the so-called system function.
%After appropriate discretization, the particle concentration can be reconstructed by solving a linear system of equations \cite{Weizenecker2007}.

In order to reconstruct an image of the MNP concentration, the physical processes during the MPI experiment need to be inverted. Mathematically, this involves the solution of an ill-posed linear system of equations \cite{Weizenecker2007}.
One prerequisite for image reconstruction is that the system matrix is accurately known. Since its very first introduction in \cite{Gleich2005}, the system matrix is measured in a data-driven calibration procedure by determining the response of the system to a small delta sample that is shifted through the imaging volume. 
The main advantage of this method is that it takes all physical effects of the imaging process into account but it has also two major drawbacks: The measured system matrix includes noise and the calibration procedure is very time consuming lasting up to several days \cite{szwargulski2018efficient}. 
These drawbacks motivated the development of methods determining the MPI system matrix based on physical models. 
%The first two milestones towards model-based MPI reconstruction where published in \cite{Knopp2009a} and \cite{Knopp2010d}. 
%In the former work it was shown that 1D MPI data could be reconstructed with high imaging quality while the latter work demonstrated successful reconstruction for 2D MPI data. 
While the initial model-based results in \cite{Knopp2009a} and \cite{Knopp2010d} were promising, model-based reconstruction based on the used {\it equilibrium model} denoting the simplified model which assumes infinitely fast relaxation leads to worse image quality than the data driven approach \cite{Kluth2017}. This is why the latter is still the method of choice in almost all publications since 2010 that use multidimensional excitation patterns. 
For 1D excitation, which includes 2D and 3D Cartesian like sampling patterns \cite{Knopp2009trajectory},%,werner2017first}, 
model-based reconstruction has been established in various works since it enables simple time-signal based reconstruction  \cite{goodwill2010x}.
%Up to now, only a limited number of works consider simplified models of the magnetic moment dynamics in the concentration reconstruction \cite{croft2012relaxation}. 

%These dynamics have not been included in multidimensional MPI image reconstruction so far due to different reasons such as for example computational complexity and unsolved parameter identification problems. 

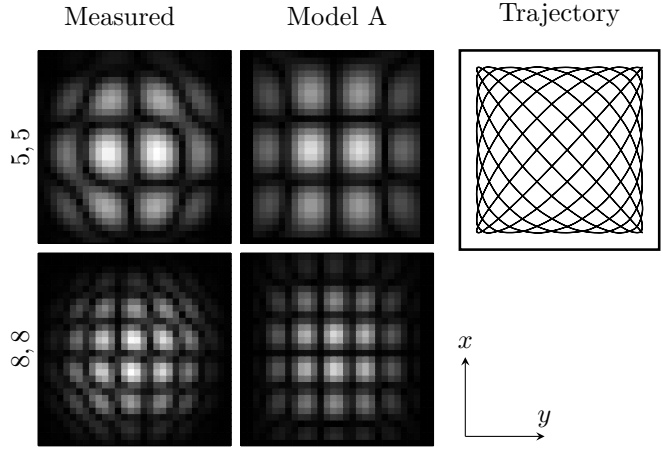
\begin{figure} 
\centering
\begin{tikzpicture}[node distance=0.01mm] 
\def\size{0.18\columnwidth}
\def\sizeshift{0.05\columnwidth}
	
	% Measured
	\node[inner sep=0.05cm](A) at (0, 0) {
	\begin{tikzpicture}
	  \begin{axis}[matrixaxis,width=\size,xticklabels=\empty,yticklabels=\empty,point meta max=1.0,point meta min=0.0, ylabel={\freqaStr},ticks=none]
            \addplot3[matrix plot*] table [x=x,y=y,z=\freqa ] {pics/SMOrig.csv};
        \end{axis}
	\end{tikzpicture}
	};
	
	\node[below= of A,inner sep=0.05cm] (A1) {
	\begin{tikzpicture}
	  \begin{axis}[matrixaxis,width=\size,xticklabels=\empty,yticklabels=\empty,point meta max=1.0,point meta min=0.0, ylabel={\freqbStr},ticks=none]
            \addplot3[matrix plot*] table [x=x,y=y,z=\freqb ] {pics/SMOrig.csv};
        \end{axis}
	\end{tikzpicture}
	};

	% Langevin
	\node[right= of A,inner sep=0.05cm] (B) {
	\begin{tikzpicture}
	  \begin{axis}[matrixaxis,width=\size,xticklabels=\empty,yticklabels=\empty,point meta max=1.0,point meta min=0.0,ticks=none]
            \addplot3[matrix plot*] table [x=x,y=y,z=\freqa ] {pics/SML.csv};
        \end{axis}
	\end{tikzpicture}
	};
	
	\node[below= of B,inner sep=0.05cm] (B1) {
	\begin{tikzpicture}
	  \begin{axis}[matrixaxis,width=\size,xticklabels=\empty,yticklabels=\empty,point meta max=1.0,point meta min=0.0,ticks=none]
            \addplot3[matrix plot*] table [x=x,y=y,z=\freqb ] {pics/SML.csv};
        \end{axis}
	\end{tikzpicture}
	};
	
	% Imob
	%\node[right= of B,inner sep=0cm] (C) {
	%\begin{tikzpicture}
	%  \begin{axis}[matrixaxis,width=\size,xticklabels=\empty,yticklabels=\empty,point meta max=1.0,point meta min=0.0,ticks=none]
    %        \addplot3[matrix plot*] table [x=x,y=y,z=\freqa ] {pics/SMImob.csv};
    %    \end{axis}
	%\end{tikzpicture}
	%};
	
	%\node[below= of C,inner sep=0cm] (C1) {
	%\begin{tikzpicture}
	%  \begin{axis}[matrixaxis,width=\size,xticklabels=\empty,yticklabels=\empty,%point meta max=1.0,point meta min=0.0,ticks=none]
    %        \addplot3[matrix plot*] table [x=x,y=y,z=\freqb ] {pics/SMImob.csv};
    %    \end{axis}
	%\end{tikzpicture}
	%};

	% Lissajous
	\node[right= of B,yshift=-0.04cm] (C) {
	\resizebox {\size} {!} {
	\begin{tikzpicture}[scale=0.9]
		\tikzstyle{ann} = [fill=white,inner sep=1.5pt]
		\draw[black,thick] (-1.2,-1.2) rectangle (1.2,1.2);
		\draw[color=black,domain=0:2.01*pi, smooth, samples=250] plot ({cos(16*\x r},{cos(17*\x r)});
	\end{tikzpicture}
	}
	};

	%labels above trajectories
	\node[above= of C, yshift=0.6mm] {Trajectory};
	\node[above= of A, yshift=1.5mm] {Measured};
	\node[above= of B, yshift=1.5mm] {Model A};
	%\node[above= of C, yshift=3mm] {Immobilized};
	
\draw[-stealth] ({2*\size-1.35cm},-3.85) -- +(30pt,0)node[above]{$y$};
\draw[-stealth] ({2*\size-1.35cm},-3.85) -- +(0,30pt)node[above]{$x$};
\end{tikzpicture}
\vspace{-0.3cm}
\caption{Imperfections of the equilibrium model compared to measured data. Absolute of two selected frequency components (denoted by mixing factors) are  shown in the first and second row. The measured system matrix using the unmodified particle solution is shown in the first column while the simulated matrix based on the equilibrium model (model A) is shown in the second column. 
The sampling pattern is shown in the last column. }
\label{fig:problem}
\end{figure}

Within this work we investigate the question why the simple equilibrium model fails to accurately describe the MPI system matrix for 2D Lissajous type excitation patterns. 
%To the best of our knowledge this is the first time considering particle dynamics in multi-dimensional excitations in the context of imaging.
As a motivating example we consider a system matrix measured along a 2D Lissajous trajectory (frequency ratio 16/17, details are outlined in Section F). 
Fig.~\ref{fig:problem} shows two selected rows of the 2D MPI system matrix and compares the measurement with the equilibrium-based simulation. 
One can see that the simulation matches quite well the basic structure of the matrix rows, which are consisting of wave pattern that share high similarity to 2D tensor products of Chebyshev polynomials \cite{Rahmer2009}. 
However, when taking a closer look one can identify various differences that lead to larger numerical deviations. %(quantitative measures are outlined in Section d). 
In particular we want to highlight two qualitative effects. 
First, the wave hills are merging in outer regions of the field-of-view (FOV), i.e. in regions where the simulation has a zero-crossing there is a non-zero value in the measured one. 
Second, one can see a tilting of the outer wave structures in outer regions of the FoV. 
The goal of this work is to simulate these effects using a physical model that takes N\'{e}el rotation of the particles' magnetic moments into account.

\subsubsection*{B. Methods} %\paragraph{Methods} \label{sec:methods}
We consider a general MPI imaging experiment where the particles are located in the MPI scanner and the change of the particle magnetization is measured using induction coils. Then the signal induced in a receive coil with sensitivity $p: \R^3 \rightarrow \R^3$ is given by
\begin{align}
 \tilde{v} (t) &=- \mu_0 \int_\Omega  c(x) p(x)^T\frac{\partial}{\partial t}\bar{m}(x,t) \ dx  \notag \label{eq:complete-problem}
\end{align}
where $c: \Omega \rightarrow \R^+_0$ is the concentration of the magnetic nanoparticles, $\bar{m}: \R^3 \times [0,T] \rightarrow \R^3$ is the mean magnetic moment of particles and $\Omega\subset\R^3$ is the imaging volume. 
The mean magnetic moment $\bar{m}(x,t)$ depends on the applied magnetic field $H_\text{app}: \R^3 \times [0,T] \rightarrow \R^3$, which is usually a $T$-periodic function with period $T$ along the time dimension. The induced signal is then convolved with the $T$-periodic analog filter $a:\R \rightarrow \R$ to remove the main contributions of the direct feedthrough yielding the signal $v=a\ast\tilde{v}$. Therefore, $v$ is $T$-periodic as well and can be expanded into a Fourier series with coefficients
\begin{align}
 \hat{v}_k &= - \hat{a}_k \frac{\mu_0}{T} \int_\Omega  c(x) p(x)^T \int_{0}^{T}  \left( \frac{\partial}{\partial t}\bar{m}(x,t) \right){\textrm{e}}^{-2\pi i tk/T} \ d t \ dx \notag
\end{align}
for $k\in\N_0$ and $\hat{v}_{-k}=\overline{\hat{v}_k}$.
This formulation is commonly used, since the signal at the excitation frequencies is blocked using an analog band-stop filter prior to the signal digitization. With $s_k(x) := - \hat{a}_k\frac{\mu_0}{T} \int_{0}^{T} p(x)^T\frac{\partial}{\partial t}\bar{m}(x,t) {\textrm{e}}^{-2\pi i tk/T} \ d t$
we can bring this into the standard notation
\begin{align}
 \hat{v}_k &=\int_\Omega  c(x) s_k(x) \ dx,  \notag
\end{align}
where $s_k: \R^3 \rightarrow \C$ is the system function that we already discussed in Fig.~\ref{fig:problem}. 
When sampling the system function $s_k$ at discrete positions $x_l, l=1,\dots,N$, and with a certain sampling rate in time one obtains the MPI system matrix $S = \left( s_k(x_l) \right)_{k=0,\dots,K; l=1,\dots,N}$. In this work we will consider both modeled system matrices $S^\text{Mod}$ as well as measures system matrices $S^\text{Cal}$ that are obtained by moving a small delta sample through the FoV. 
In case of multiple receive channels, the corresponding system matrices can be stacked to arrive at a joint linear system of equation; see for example \cite{Kluth2019numerical} for a formal definition.

%{\color{blue} [TKl: Hier kurz das Fitten der TF anfuehren? 1-2 Sätze plus Literaturverweis fuer Details. Sollte einmal knapp erwaehnt werden.]}
\subsubsection*{C. Particle Models} %\paragraph{Particle Models} \label{sec:models}
Since $H_\text{app}$ and $p$ can be easily modeled using the Biot-Savart law and $a$ can either measured or fitted \cite{Knopp2010d}, we focus on modeling the particles' mean magnetic moment behavior. 
We first consider the equilibrium model that can be derived under the assumption that the particles are always in thermal equilibrium in which case $\bar{m}$ directly follows the applied field $H_\text{app}$. 
We refer to it as {\it model A} which is given by
\begin{equation}
 \bar{m}(x,t) = m_0 \mathcal{L}_\beta(| H_\mathrm{app}(x,t)|) \frac{H_\mathrm{app}(x,t)}{| H_\mathrm{app}(x,t)|} 
\end{equation}
where $\mathcal{L}_\beta: \R \rightarrow \R$ is given in terms of the Langevin function by the following:
\begin{equation}
\label{eq:Langevin}
 \mathcal{L}_\beta(z) =  \left( \coth( \beta z) - \frac{1}{ \beta z} \right)
\end{equation}
for $m_0,\beta >0$. Both $m_0=V_\mathrm{C} M_\mathrm{S}$ and $\beta=\frac{\mu_0 V_\mathrm{C} M_\mathrm{S}}{k_\mathrm{B} T_\mathrm{B}}$ mainly depend on the particle core diameter $D$.

Since we have already seen that the equilibrium model is not accurate enough for modeling the physics in an MPI experiments we next consider a more appropriate model that takes dynamic relaxation effects into account. There are basically two different approaches, one can either consider the problem on a micromagnetic level for individual particles and solve the Langevin equation 
corresponding to the Landau-Lifschitz-Gilbert equation for a sufficiently large number of particles to obtain a reasonable estimate for the mean. Alternatively, one can take a comprehensive view and solve the Fokker-Planck equation for a probability distribution representing an entire ensemble of nanoparticles in terms of a parabolic partial differential equation. The latter approach benefits from efficient and more accurate solutions when aiming for a mean computation \cite{Weizenecker2018} as determining the mean from the Langevin equation requires a large number of particle simulations. 

The second case which we refer to as {\it model B} is based on the latter approach which determines the mean magnetic moment via the probability density function $f: S^2 \times \Omega \times [0,T] \rightarrow \R^+\cup \{0\}$ which is the solution to the corresponding Fokker-Planck equation where $S^2$ is the surface of the sphere in $\R^3$.
The mean is then given by
\begin{equation}
  \bar{m}(x,t) = m_0 \int_{S^2} m f(m,x,t) \ dm
\end{equation}
where $f$ is the solution to the following specific case of a convection-diffusion equation on the sphere
\begin{equation}
\label{eq:FP-general}
 \frac{\partial}{\partial t} f = \mathrm{div}_{S^2}(\frac{1}{2\tau} \nabla_{S^2} f ) - \mathrm{div}_{S^2}(a f)
\end{equation}
where $\tau >0$ is the relaxation time constant and the (velocity) field $a:S^2 \times \R^3 \times S^2 \rightarrow \R^3$ given by
\begin{align}
 &a(m,H_\mathrm{app},n) = p_1 H_\mathrm{app} \times m + p_2 (m\times H_\mathrm{app}) \times m \notag \\ & \quad+ p_3 (n,m) n \times m + p_4 (n,m) (m\times n) \times m \label{eq:convection}
\end{align}
where $p_i\geq 0$, $i=1,\hdots,4$, and $n\in S^2$ is the easy axis of the particles (or alternatively $n:\Omega \rightarrow S^2$ and $p_3,p_4: \Omega \rightarrow \R_0^+$). 
Differentiation in terms of gradient $\nabla_{S^2} $ and divergence $ \mathrm{div}_{S^2}$ is considered with respect to the surface $S^2$.
$(\cdot,\cdot)$ denotes the Euclidean scalar product of $\R^3$.
The specific structure of the Fokker-Planck equation is due to the independent consideration of Brownian and N\'{e}el rotation model as it was observed for common tracer that one meachism can be dominant \cite{ludwig2013characterization}.
  For a detailed discussion of the relationship between Fokker-Planck equation and Langevin equations for Brownian and N\'{e}el rotation we refer to the survey \cite{Kluth2018a}.
 A pure N\'{e}el rotation including anisotropy is given by $p_1=\tilde{\gamma}\mu_0$, $p_2=\tilde{\gamma}\alpha \mu_0$, $p_3=2\tilde{\gamma}\frac{K_\mathrm{anis}}{M_\mathrm{S}}$, $p_4=2\tilde{\gamma}\frac{K_\mathrm{anis}}{M_\mathrm{S}}$, and $\tau=\frac{V_\mathrm{C} M_\mathrm{S}}{2 k_\mathrm{B} T_\mathrm{B} \tilde{\gamma} \alpha}$ ($\tilde{\gamma}=\frac{\gamma}{1+ \alpha^2}$).
%\begin{remark}
 We note that the parabolic PDE in \myEqref{eq:FP-general} has no dependence on derivatives with respect to the spatial variable $x$.
 It can thus be considered as parametric with respect to $x$.
%\end{remark}

In the following we use the N\'{e}el rotation model as it includes the particle anisotropy. 
Here we distinguish the following three cases to identify the relevant modeling aspects:
\begin{itemize}
 \item[B1] N\'{e}el rotation model without anisotropy, i.e., $p_3=p_4=0$ in \myEqref{eq:convection}.
 \item[B2] N\'{e}el rotation model including anisotropy, i.e., it includes all summands in \myEqref{eq:convection} for a given easy axis $n\in S^2$.
 \item[B3] N\'{e}el rotation model including a space-dependent anisotropy which can be motivated by the local structure of the magnetic field, i.e., $n:\Omega \rightarrow S^2$. 
 In particular in the corners of the FoV the magnetic field vector is mainly oriented within one quadrant. 
 During multiple repetitions of the excitation sequence this can lead to a preferred orientation of the easy axis of the nanoparticles in a ferrofluid, which can highly influence the magnetization behavior as reported in \cite{Yoshida2017}  
\end{itemize}

%\begin{remark}
 %Note that different choices of parameter settings for $p_i$ cover the cases of interest. A pure N\'{e}el rotation including anisotropy is given by $p_1=$, $p_2=$, $p_3=$, $p_4=$, and $tau=$.
% The Brownian case is covered by the parameter set $p_2=\mu_0\frac{V_\mathrm{C} M_\mathrm{S}}{6 \eta V_\mathrm{H}}$, $p_1=p_3=p_4=0$, and $\tau=\frac{3 V_\mathrm{H}\eta}{k_\mathrm{B} T_\mathrm{B}}$. 
%\end{remark}

Following the approach in \cite{Weizenecker2018} the numerical solution of \myEqref{eq:FP-general} is obtained as follows: 
We consider an initial value $f(m,x,0)=f_0(m,x)$ with $f_0(\cdot,x)\geq 0$  and $\|f_0(\cdot,x)\|_{L^1(S^2)}=1$ for any $x\in\Omega$.
A semi-discrete Cauchy problem is obtained via discretization in non-normalized spherical harmonics $\{Y_l^{m'}\}_{l,m'}$ (with respect to $m$ while considering a rotated problem using a rotation matrix $R\in \R^{3\times 3}$ such that $Rn=e_3$) which yields a first order ordinary differential equation system for any $x\in \Omega$. 
%For a more detailed consideration of the discretization in line with the approach in \cite{Weizenecker2018} we refer to Appendix \ref{sec:discretizationFP}.
The initial value problem is then solved on a time interval covering multiple periods of the excitation pattern while using a uniform initial value (for any $x\in\Omega$).
The numerical solution is obtained via a variable order, variable step solver using Matlab's builtin function \texttt{ode15s}.
The last period of the solution is then used for further investigations.
The mean magnetic moment can then be determined directly via linear combinations of the coefficients corresponding to the spherical harmonics $Y_1^0$, $Y_1^{1}$, and $Y_1^{-1}$, which is subsequently rotated back. % to the original problem. 

\subsubsection*{D. Distance Measure} 
%\paragraph{Distance measure} \label{sec:distance_measures}
For quantifying the accuracy of a model 
we express the frequency index $k$ in terms of the mixing orders $m_x$, $m_y$ \cite{rahmer2012analysis} fulfilling
% an appropriate error metric is needed. While in principle it would be possible to use measures that are based on matrix norms this does not take the special structure of the MPI system matrix into account. We therefore use a custom metric that is based on the mixing factors. As discussed in \cite{rahmer2012analysis} the row index $k$ can be expressed in terms of the factors $m_x$ and $m_y$ fulfilling
\begin{align} \label{Eq:mix}
    k(m_x,m_y) &= m_x N_{\text{Dens}} + m_y (N_{\text{Dens}}+1)
\end{align}
Here, $N_{\text{Dens}}$  is the parameter that controls the density of the Lissajous sampling pattern and for each $k$ the factors $m_x$ and $m_y$ with the smallest absolute value fulfilling \eqref{Eq:mix} are considered.
Now with $S^\text{Cal}_{k(m_x,m_y)}$ being the measured and $S^\text{Mod}_{k(m_x,m_y)}$ being the modeled system matrix row we consider the the normalized root mean square error (NRMSE)
\begin{align} \label{Eq:error1}
    \varepsilon(S^\text{Cal}_k, S^\text{Mod}_k) &= \frac{\frac{1}{\sqrt{N}}\Vert  S^\text{Cal}_k- S^\text{Mod}_k \Vert_2}{\Vert  S^\text{Cal}_k \Vert_\infty}.
\end{align}
In addition to this per-row metric we consider the mean NRMSE over the $m_x=0,\dots,M_x$ and $m_y=0,\dots,M_y$ where $M_x,M_y\in\N$ are upper bounds for the mixing order.
%By applying the normalization, the metric is independent of the norm of the system matrix rows, which can vary drastically for the MPI system matrix. Based on this per row difference metric we define a global metric (mean NRMSE)
%\begin{align*} \label{Eq:error2}
%    %&\varepsilon^\text{g}(S^\text{Cal}, S^\text{Mod}) =
%    \frac{1}{M_x M_y} \sum_{m_x=0}^{M_x}\sum_{m_y=0}^{M_y} \varepsilon(S^\text{Cal}_{k(m_x,m_y)}, S^\text{Mod}_{k(m_x,m_y)})
%\end{align*}
%that depends on $M_x,M_y\in\N$.

\subsubsection*{E. Image Reconstruction} %\paragraph{Image reconstruction}
For image reconstruction we use the standard framework \cite{knopp2016online} that applies the iterative Kaczmarz algorithm augmented by Tikhonov regularization and a positivity constraint. We apply standard SNR thresholding where matrix rows with low SNR (< 2.0) are removed prior to reconstruction.
In addition we apply a second way of matrix row filtering. The idea is to select only those matrix rows of the modeled system matrices, which deviate only marginal from the measured one. Let $\Theta \in [0,1]$ be a predefined threshold. Then, we choose matrix rows $k$ where $\varepsilon(S^\text{Cal}_{k}, S^\text{Mod}_{k}) < \Theta$. The parameter $\Theta$ can now be used to trade of between inaccuracies due to the model/measurement mismatch and a loss off spatial resolution that happens when only few system matrix rows are selected for reconstruction.

\subsubsection*{F. Experimental Setup}
%\paragraph{Experimental setup} \label{sec:experimentalsetup}
Experimental data was measured with a preclinical MPI scanner (Bruker, Ettlingen) using field-free-point (FFP) excitation patterns. In all experiments a 2D cosine excitation ($N_{\text{Dens}}=16$) within the $xy$-plane of the scanner was applied with excitation-field amplitudes $A_x=A_y=$ 12~mT$\mu_0^{-1}$ and 1~Tm$^{-1}\mu_0^{-1}$ selection field gradient within the $xy$-plane. The resulting area covered by the FFP is of size $24\times 24$~mm$^2$.

\begin{figure} 
\centering
\input{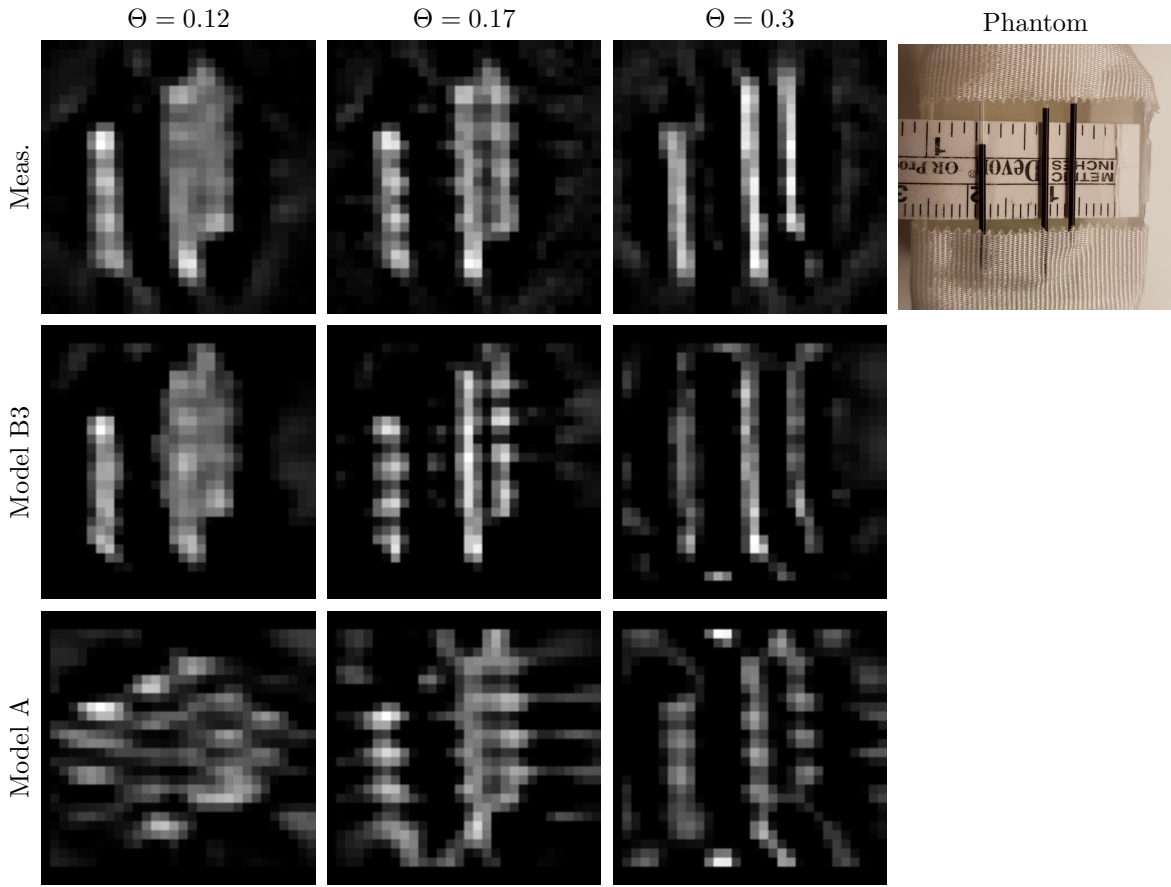}
\vspace{-0.3cm}
\caption{Reconstruction results of measured phantom data using three different system matrices: 
The measured system matrix, the matrix based on the equilibrium model A, and the system matrix obtained from the N\'{e}el model B3 for $g^3_{K_\text{anis}}=1250/h~\text{J/m$^3$}$. 
Each reconstruction uses the same regularization parameter $\lambda=10^{-7}$ and the threshold $\Theta$ is varied between $0.12$ and $0.3$.
%{\color{blue}[TKl: Ist der jeweilige lineare Operator normalisiert? Wieviel Iterationen? Zusätzliches SNR Thresholding angewandt? Oder Details aus Platzgruenden nicht aufnehmen?] }
}
\label{fig:reco}
\end{figure}

The system matrix was measured by shifting a 1~mm$^3$ delta sample of Perimag (Micromod, Rostock) to $30\times 30$ positions covering a FoV of size $30\times 30$~mm$^2$ (1~mm step size). 
%One delta sample was measured with unmodified perimag (micromod, Rostock), while the other was measured with particles immobilized using tooth cement.
A particle phantom consisting of three rods 1.0~mm in diameter is prepared using the same tracer of the same concentration as the delta sample. The phantom is shown in Fig.~\ref{fig:reco}.

%[TKl: Hier muesste noch das Setup genauer beschrieben werden. Die immobilisierten Messungen benoetigen wir an dieser Stelle bisher nicht; dann erst in der langen Studie mit aufnehmen, richtig? Wie groß war die Delta-Probe (Bzgl. der getrockneten Probe meinte Florian am Telefon 3x3x1.5mm)? Step length fuer Positionen war 1 mm laut mdf Datei, oder?]

\subsubsection*{G. Model Parameter Selection}
%\paragraph{Model parameter selection}
The considered models each have several parameters that are a-priori unknown. 
Since the simulation of the models requires a significant time, it is challenging to apply for example gradient-based optimization techniques to the continuous parameter space. 
Therefore, we calculate system matrices for a finite number of settings and perform a subsequent discrete optimization.

Model A as well as models B1-B3 require as input the particle core diameter $D$. 
We simulated the equilibrium model for $D\in \{20~\text{nm},25~\text{nm},30~\text{nm}\}$ and found in initial tests that the system matrices with $D=20~\text{nm}$ fitted best in all model cases and are therefore considered for further investigation. 
For the N\'{e}el model B2 we consider a spatially homogeneous easy axis distribution $n(\theta)=(\cos(\theta),\sin(\theta))^T$ and orientations $\theta \in \{0^\circ, 45^\circ, 90^\circ,135^\circ \}$. 
In this case the anisotropy constant is chosen as $K_\text{anis}=625~\text{J/m$^3$}$.

For the inhomogeneous case B3 we consider an easy axis $n(x)=\frac{x}{\vert x \vert}$ and $K_\mathrm{anis}(x)=g_{K_\mathrm{anis}} | H_\mathrm{S}(x) |$ where we chose $g_{K_\mathrm{anis}}  \in \{ 625/h, 1250/h, 2500/h, 5000/h \} \text{J/m$^3$}$ with $h=\max_{x\in\Omega} |H_\mathrm{S}(x)|$.
Thus, the easy axis is parallel to the selection field vector and the effective anisotropy increases when increasing the distance to the FFP of the selection field.

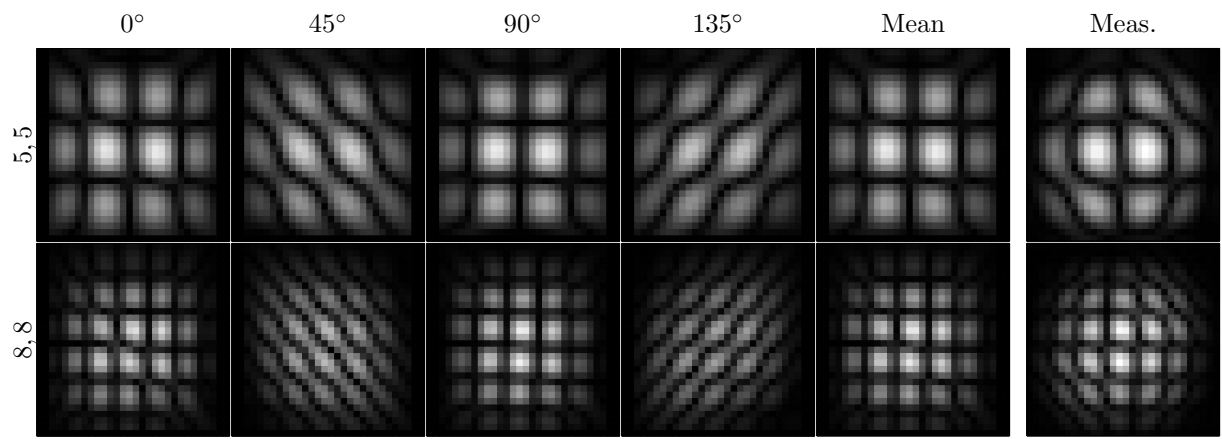
\begin{figure} 
\centering
\begin{tikzpicture}[node distance=0.01mm] 
\def\size{0.18\columnwidth}
\def\sizeshift{0.1\columnwidth}
	
	\node[inner sep=0cm](A) at (0, 0) {
	\begin{tikzpicture}
	  \begin{axis}[matrixaxis,width=\size,xticklabels=\empty,yticklabels=\empty,point meta max=1.0,point meta min=0.0, title={$0^\circ$}, ylabel={\freqaStr}, ticks=none]
            \addplot3[matrix plot*] table [x=x,y=y,z=\freqa ] {pics/SMD1Easy1Anis3.csv};
        \end{axis}
	\end{tikzpicture}
	};
	
	\node[below= of A,inner sep=0cm] (A1) {
	\begin{tikzpicture}
	  \begin{axis}[matrixaxis,width=\size,xticklabels=\empty,yticklabels=\empty,point meta max=1.0,point meta min=0.0, ylabel={\freqbStr}, ticks=none]
            \addplot3[matrix plot*] table [x=x,y=y,z=\freqb ] {pics/SMD1Easy1Anis3.csv};
        \end{axis}
	\end{tikzpicture}
	};
	
	\node[right= of A,inner sep=0cm] (B) {
	\begin{tikzpicture}
	  \begin{axis}[matrixaxis,width=\size,xticklabels=\empty,yticklabels=\empty,point meta max=1.0,point meta min=0.0, title={$45^\circ$}, ticks=none]
            \addplot3[matrix plot*] table [x=x,y=y,z=\freqa ] {pics/SMD1Easy3Anis3.csv};
        \end{axis}
	\end{tikzpicture}
	};
	
	\node[below= of B,inner sep=0cm] (B1) {
	\begin{tikzpicture}
	  \begin{axis}[matrixaxis,width=\size,xticklabels=\empty,yticklabels=\empty,point meta max=1.0,point meta min=0.0, ticks=none]
            \addplot3[matrix plot*] table [x=x,y=y,z=\freqb ] {pics/SMD1Easy3Anis3.csv};
        \end{axis}
	\end{tikzpicture}
	};
	
	\node[right= of B,inner sep=0cm] (C) {
	\begin{tikzpicture}
	  \begin{axis}[matrixaxis,width=\size,xticklabels=\empty,yticklabels=\empty,point meta max=1.0,point meta min=0.0, title={$90^\circ$}, ticks=none]
            \addplot3[matrix plot*] table [x=x,y=y,z=\freqa ] {pics/SMD1Easy2Anis3.csv};
        \end{axis}
	\end{tikzpicture}
	};
	
	\node[below= of C,inner sep=0cm] (C1) {
	\begin{tikzpicture}
	  \begin{axis}[matrixaxis,width=\size,xticklabels=\empty,yticklabels=\empty,point meta max=1.0,point meta min=0.0, ticks=none]
            \addplot3[matrix plot*] table [x=x,y=y,z=\freqb ] {pics/SMD1Easy2Anis3.csv};
        \end{axis}
	\end{tikzpicture}
	};

	\node[right= of C,inner sep=0cm] (D) {
	\begin{tikzpicture}
	  \begin{axis}[matrixaxis,width=\size,xticklabels=\empty,yticklabels=\empty,point meta max=1.0,point meta min=0.0, title={$135^\circ$}, ticks=none]
            \addplot3[matrix plot*] table [x=x,y=y,z=\freqa ] {pics/SMD1Easy4Anis3.csv};
        \end{axis}
	\end{tikzpicture}
	};
	
	\node[below= of D,inner sep=0cm] (D1) {
	\begin{tikzpicture}
	  \begin{axis}[matrixaxis,width=\size,xticklabels=\empty,yticklabels=\empty,point meta max=1.0,point meta min=0.0, ticks=none]
            \addplot3[matrix plot*] table [x=x,y=y,z=\freqb ] {pics/SMD1Easy4Anis3.csv};
        \end{axis}
	\end{tikzpicture}
	};

	\node[right= of D,inner sep=0cm] (E) {
	\begin{tikzpicture}
	  \begin{axis}[matrixaxis,width=\size,xticklabels=\empty,yticklabels=\empty,point meta max=1.0,point meta min=0.0, title={Mean}, ticks=none]
            \addplot3[matrix plot*] table [x=x,y=y,z=\freqa ] {pics/SMD1MeanEasyAnis3.csv};
        \end{axis}
	\end{tikzpicture}
	};
	
	\node[below= of E,inner sep=0cm] (E1) {
	\begin{tikzpicture}
	  \begin{axis}[matrixaxis,width=\size,xticklabels=\empty,yticklabels=\empty,point meta max=1.0,point meta min=0.0, ticks=none]
            \addplot3[matrix plot*] table [x=x,y=y,z=\freqb ] {pics/SMD1MeanEasyAnis3.csv};
        \end{axis}
	\end{tikzpicture}
	};
	
	%\node[right= of E,inner sep=0cm,xshift=2mm] (F) {
	%\begin{tikzpicture}
	%  \tikz \fill [lightgray] (0.1,1.7) rectangle ++(1.6,5.5);
	%\end{tikzpicture}
	%};	

	\node[right= of E,inner sep=0cm,xshift=2mm] (F) {
	\begin{tikzpicture}
	  \begin{axis}[matrixaxis,width=\size,xticklabels=\empty,yticklabels=\empty,point meta max=1.0,point meta min=0.0, title={Meas.}, ticks=none]
            \addplot3[matrix plot*] table [x=x,y=y,z=\freqa ] {pics/SMOrig.csv};
        \end{axis}
	\end{tikzpicture}
	};
	
	\node[below= of F,inner sep=0cm] (F1) {
	\begin{tikzpicture}
	  \begin{axis}[matrixaxis,width=\size,xticklabels=\empty,yticklabels=\empty,point meta max=1.0,point meta min=0.0, ticks=none]
            \addplot3[matrix plot*] table [x=x,y=y,z=\freqb ] {pics/SMOrig.csv};
        \end{axis}
	\end{tikzpicture}
	};
	
\end{tikzpicture}
\vspace{-0.3cm}
\caption{Selected frequency components of modeled system matrices using the N\'{e}el model with anisotropy constant $625~\text{J/m$^3$}$ and a homogeneous easy axis for 4 different orientations (columns $1-4$). 
Column 5 shows the mean over all easy axes while column 6 shows the measured system matrix for reference.}
\label{fig:easy}
\end{figure}

\begin{figure} 
\centering
\begin{tikzpicture}[node distance=0.01mm] 
\def\size{0.18\columnwidth}
\def\sizeshift{0.1\columnwidth}
	
	\node[inner sep=0cm](A) at (0, 0) {
	\begin{tikzpicture}
	  \begin{axis}[matrixaxis,width=\size,xticklabels=\empty,yticklabels=\empty,point meta max=1.0,point meta min=0.0, ylabel={\freqaStr}, ticks=none]
            \addplot3[matrix plot*] table [x=x,y=y,z=\freqa ] {pics/SMD1Anis1.csv};
        \end{axis}
	\end{tikzpicture}
	};
	\node[above= of A, yshift=2.1mm] {$g_{K_\text{Anis}}^1$};
	
	\node[below= of A,inner sep=0cm] (A1) {
	\begin{tikzpicture}
	  \begin{axis}[matrixaxis,width=\size,xticklabels=\empty,yticklabels=\empty,point meta max=1.0,point meta min=0.0, ylabel={\freqbStr}, ticks=none]
            \addplot3[matrix plot*] table [x=x,y=y,z=\freqb ] {pics/SMD1Anis1.csv};
        \end{axis}
	\end{tikzpicture}
	};
	
	\node[right= of A,inner sep=0cm] (B) {
	\begin{tikzpicture}
	  \begin{axis}[matrixaxis,width=\size,xticklabels=\empty,yticklabels=\empty,point meta max=1.0,point meta min=0.0, ticks=none]
            \addplot3[matrix plot*] table [x=x,y=y,z=\freqa ] {pics/SMD1Anis2.csv};
        \end{axis}
	\end{tikzpicture}
	};
	\node[above= of B, yshift=2.1mm] {$g_{K_\text{Anis}}^2$};
	
	\node[below= of B,inner sep=0cm] (B1) {
	\begin{tikzpicture}
	  \begin{axis}[matrixaxis,width=\size,xticklabels=\empty,yticklabels=\empty,point meta max=1.0,point meta min=0.0, ticks=none]
            \addplot3[matrix plot*] table [x=x,y=y,z=\freqb ] {pics/SMD1Anis2.csv};
        \end{axis}
	\end{tikzpicture}
	};
	
	\node[right= of B,inner sep=0cm] (C) {
	\begin{tikzpicture}
	  \begin{axis}[matrixaxis,width=\size,xticklabels=\empty,yticklabels=\empty,point meta max=1.0,point meta min=0.0, ticks=none]
            \addplot3[matrix plot*] table [x=x,y=y,z=\freqa ] {pics/SMD1Anis3.csv};
        \end{axis}
	\end{tikzpicture}
	};
	\node[above= of C, yshift=2.1mm] {$g_{K_\text{Anis}}^3$};
	
	\node[below= of C,inner sep=0cm] (C1) {
	\begin{tikzpicture}
	  \begin{axis}[matrixaxis,width=\size,xticklabels=\empty,yticklabels=\empty,point meta max=1.0,point meta min=0.0, ticks=none]
            \addplot3[matrix plot*] table [x=x,y=y,z=\freqb ] {pics/SMD1Anis3.csv};
        \end{axis}
	\end{tikzpicture}
	};

	\node[right= of C,inner sep=0cm] (D) {
	\begin{tikzpicture}
	  \begin{axis}[matrixaxis,width=\size,xticklabels=\empty,yticklabels=\empty,point meta max=1.0,point meta min=0.0, ticks=none]
            \addplot3[matrix plot*] table [x=x,y=y,z=\freqa ] {pics/SMD1Anis4.csv};
        \end{axis}
	\end{tikzpicture}
	};
	\node[above= of D, yshift=2.1mm] {$g_{K_\text{Anis}}^4$};
	
	\node[below= of D,inner sep=0cm] (D1) {
	\begin{tikzpicture}
	  \begin{axis}[matrixaxis,width=\size,xticklabels=\empty,yticklabels=\empty,point meta max=1.0,point meta min=0.0, ticks=none]
            \addplot3[matrix plot*] table [x=x,y=y,z=\freqb ] {pics/SMD1Anis4.csv};
        \end{axis}
	\end{tikzpicture}
	};
	
	\node[right= of D,inner sep=0cm,xshift=1mm] (E) {
	\begin{tikzpicture}
	  \begin{axis}[matrixaxis,width=\size,xticklabels=\empty,yticklabels=\empty,point meta max=1.0,point meta min=0.0, ticks=none]
            \addplot3[matrix plot*] table [x=x,y=y,z=\freqa ] {pics/SMD1NoAnis.csv};
        \end{axis}
	\end{tikzpicture}
	};
	\node[above= of E, yshift=2.1mm] {Model B1};
	
	\node[below= of E,inner sep=0cm] (E1) {
	\begin{tikzpicture}
	  \begin{axis}[matrixaxis,width=\size,xticklabels=\empty,yticklabels=\empty,point meta max=1.0,point meta min=0.0, ticks=none]
            \addplot3[matrix plot*] table [x=x,y=y,z=\freqb ] {pics/SMD1NoAnis.csv};
        \end{axis}
	\end{tikzpicture}
	};
	
	%\node[right= of D,inner sep=0cm,xshift=2mm] (U) {
	%\begin{tikzpicture}
	%  \tikz \fill [lightgray] (0.1,0.8) rectangle ++(1.9,3.6);
	%\end{tikzpicture}
	%};	

	\node[right= of E,inner sep=0cm,xshift=1mm] (F) {

	\begin{tikzpicture}
	  \begin{axis}[matrixaxis,width=\size,xticklabels=\empty,yticklabels=\empty,point meta max=1.0,point meta min=0.0, ticks=none]
            \addplot3[matrix plot*] table [x=x,y=y,z=\freqa ] {pics/SMOrig.csv};
        \end{axis}
	\end{tikzpicture}
	};
	\node[above= of F, yshift=2.1mm] {Meas.};
	
	\node[below= of F,inner sep=0cm] (F1) {
	\begin{tikzpicture}
	  \begin{axis}[matrixaxis,width=\size,xticklabels=\empty,yticklabels=\empty,point meta max=1.0,point meta min=0.0, ticks=none]
            \addplot3[matrix plot*] table [x=x,y=y,z=\freqb ] {pics/SMOrig.csv};
        \end{axis}
	\end{tikzpicture}
	};
	
	\node[below= of A1,inner sep=0.1cm,xshift=6.5cm] (B) {
	\begin{tikzpicture}
	  \pgfplotsset{every axis/.append style={line width=1pt},
                                       cycle list name = exotic,
                                       width=0.99*\columnwidth,
                                       height=0.45*\columnwidth
                                      }
		\begin{axis}[legend,ymax=0.195,ymin=0.10,xmin=1,xmax=4,ylabel=mean NRMSE,xlabel={Anisotropy gradient}, legend style={at={(0.6,0.85)},anchor=west}]
		\addplot[black!40,line width=3pt,thick] coordinates {
		(1,0.1505)
		(4,0.1505)
	    }; 
		\addlegendentry{Model A};
		\addplot[black!70,line width=1pt,dashed] coordinates {
		(1,0.13408114559058099)
		(4,0.13408114559058099)
	    }; 
		\addlegendentry{Model B1};
		\addplot[black,line width=3pt,thick,mark=*] coordinates {
		(1,0.19346124172958032)
		(2,0.1366518358509561)
		(3,0.1111006875842072)
		(4,0.11525709377341005)
	    }; 
	    \addlegendentry{Model B3};
		\end{axis}
	\end{tikzpicture}
	};
\end{tikzpicture}
\vspace{-0.7cm}
\caption{Selected frequency components of modeled system matrices using the N\'{e}el model B3 with anisotropy gradients ranging from $5000/h~\text{J/m$^3$}$ to $625/h~\text{J/m$^3$}$ (columns 1-4;left to right). Column 5 shows the modeled system matrix using the N\'{e}el model B1 without anisotropy while column 6 shows the measured system matrix for reference. 
Below, the mean NRMSE between the model and the measurement is shown.} 
\label{fig:anis}
\end{figure}

\begin{figure}[t!]
\centering
\includegraphics[width=0.99\columnwidth]{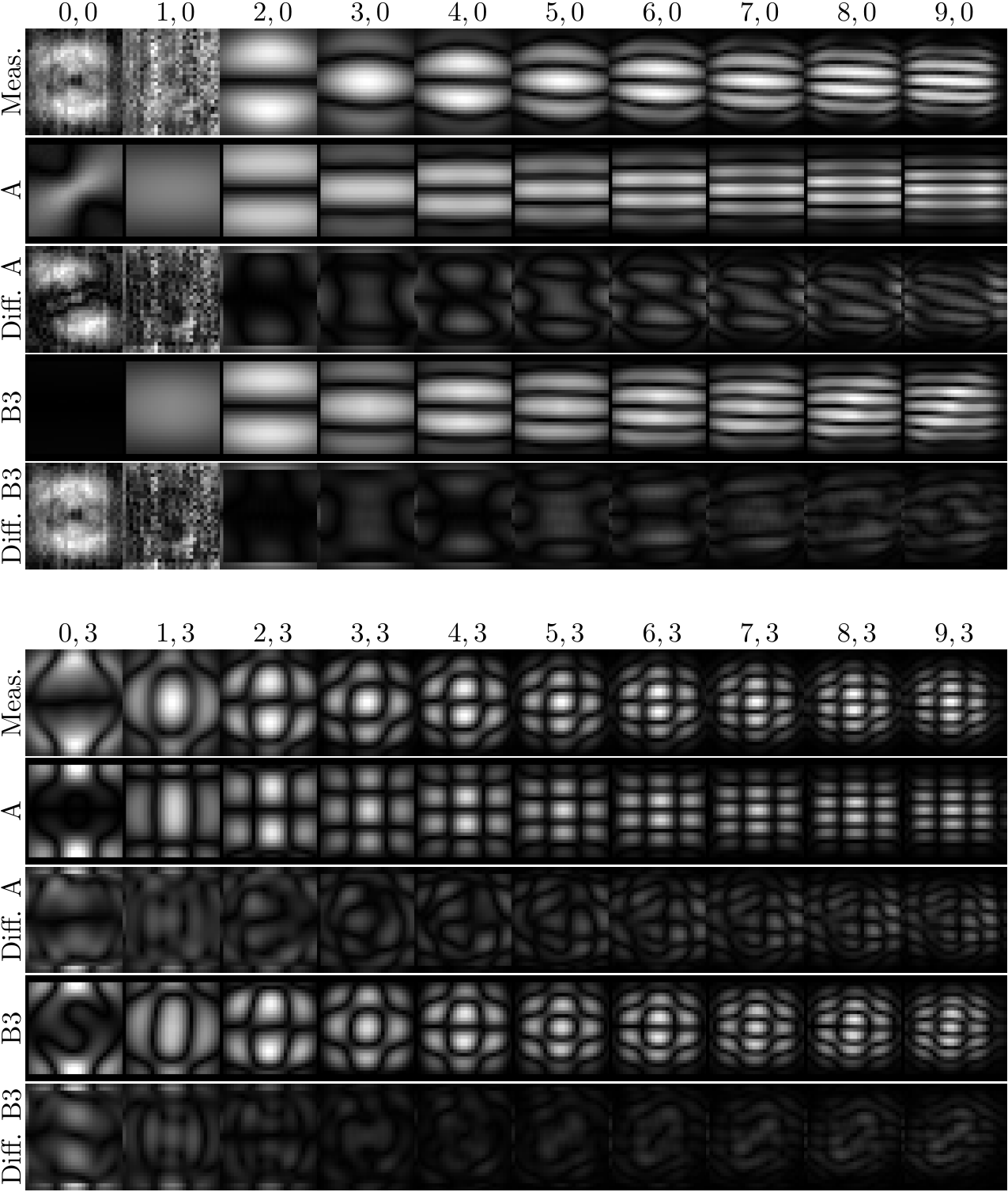}
\caption{Comparison of measured and modeled system matrices for the equilibrium model A and the N\'{e}el model B3 for $g^3_{K_\text{anis}}=1250/h~\text{J/m$^3$}$. 
Shown are frequency components for combinations of $m_x=0,\dots, 9$ and $m_y=0,4$. 
Below in the ``Diff.'' rows of the respective model, the absolute difference between measurement and model is shown. 
}
 \label{Fig:SMComp}
\end{figure}

\begin{figure} 
\centering
\begin{tikzpicture}[node distance=0.01mm] 
\def\size{0.3\columnwidth}
\def\sizeshift{0.05\columnwidth}
	
	\node[inner sep=0.1cm](A) at (0, 0) {
	\begin{tikzpicture}
	  \begin{axis}[matrixaxis,width=\size,point meta max=0.35,point meta min=0.05, title={Model A}, xlabel={$m_x$}, ylabel={$m_y$}]
            \addplot3[matrix plot*] table [x=x,y=y,z=error] {pics/ErrorL.csv};
        \end{axis}
	\end{tikzpicture}
	};
	
	\node[right= of A,inner sep=0.1cm] (B) {
	\begin{tikzpicture}
	  \begin{axis}[matrixaxis,width=\size,point meta max=0.35,point meta min=0.05, title={Model B3},colorbar,colorbar style={ytick={0.05,0.35}}, xlabel={$m_x$}, yticklabels=\empty]
            \addplot3[matrix plot*] table [x=x,y=y,z=error] {pics/ErrorD1Anis3.csv};
        \end{axis}
	\end{tikzpicture}
	};
	
	%\node[below= of A,inner sep=0.1cm,xshift=2.4cm] (B) {
	%\begin{tikzpicture}
	%  \pgfplotsset{every axis/.append style={line width=1pt},
    %                                   cycle list name = exotic,
    %                                   width=0.95*\columnwidth,
    %                                   height=0.45*\columnwidth
    %                                  }
%		\begin{axis}[legend,ymax=0.39,ymin=0.05,xmin=0,xmax=100,xtick distance=10,ylabel=NRMSE,xlabel={$m_x + 9m_y$}, legend style={at={(0.02,0.85)},anchor=west}]
%		\addplot[black!40,line width=2pt,thick,mark=*] table[x=k, y=error]{pics/ErrorL.csv};
%		\addlegendentry{Model A};
%		\addplot[black,line width=2pt,thick,mark=*] table[x=k, y=error]{pics/ErrorD1Anis3.csv};
%		\addlegendentry{Model B3};
%		\end{axis}
%	\end{tikzpicture}
%	};
\end{tikzpicture}
\vspace{-0.3cm}
\caption{Error of the modeled system matrices (A and B3) shown in Fig.~\ref{Fig:SMComp}. 
The error is shown in two individual gray scale images (top) and for direct comparison in a flattened plot. }
\label{fig:error}
\end{figure}
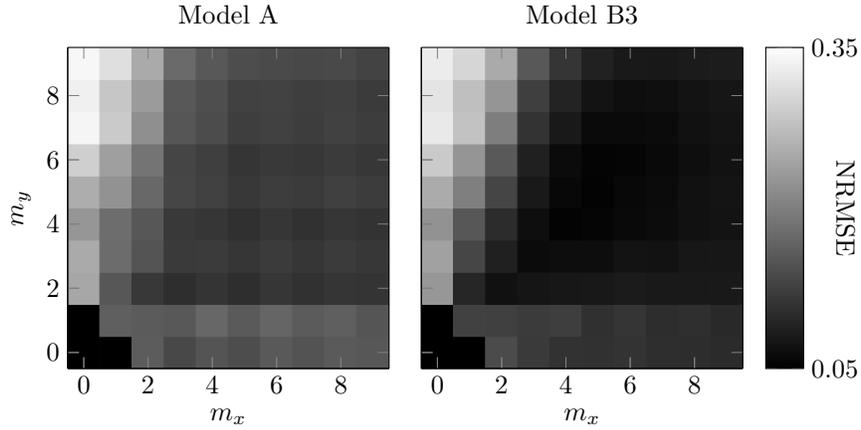

\subsubsection*{H. Results} %\paragraph{Results}
First, we consider the N\'{e}el model B2 with spatially homogeneous anisotropy distribution and homogeneous easy axis. Selected rows of simulated system matrices are shown in Fig.~\ref{fig:easy}. 
One can see that the angles $0^\circ$ and $90^\circ$ do not change the shape of the patterns compared to the equilibrium model. 
However, for $45^\circ$ and $135^\circ$ one can see the desired effect that the patterns are tilted in one direction, which is the orientation of the easy axis. 
In addition one can also see the merge of the wave hills that was observed in the measured system matrices. 
However, in direct comparison with the measured system matrix, one can see that the direction of the tilt is homogeneous while the measurement shows a spatially dependent tilt. 
One attempt could be to assume that there is a homogeneous distribution of the easy axis in each voxel, which can be simulated by taking the mean over all simulated angles. 
But as is shown in Fig.~\ref{fig:easy}, applying the mean leads to a similar pattern as observed for the equilibrium model A (see Fig.~\ref{fig:problem}).

The results from the spatially homogeneous anisotropy distribution motivate the usage of an space-dependent anisotropy distribution. 
In order to make the tilting orientation-dependent, we let the easy axis point along radial direction being parallel to the selection field vector. 
%In order to make the effect stronger in outer regions, we linear increase the anisotropy constant for different anisotropy gradients $g_{K_\text{anis}}$ ranging from $g^4_{K_\text{anis}}=625/h~\text{J/m$^3$}$ to $g^1_{K_\text{anis}}=5000/h~\text{J/m$^3$}$. 
The results are shown in Fig.~\ref{fig:anis}.  
One can see that the introduction of the spatially inhomogeneous easy axis and anisotropy constant does indeed introduce the two effects observed in the measurements. 
First a tilting effect and second a merging of the wave hills. 
With increasing anisotropy gradient, the effect gets stronger and starts already at the center. 
For small gradients, the effect is only visible in outer regions. 
For $g^3_{K_\text{anis}}=1250/h~\text{J/m$^3$}$ we find the highest similarity compared to the measurement, which is also supported by the mean NRMSE shown in Fig.~\ref{fig:anis}.

After choosing and optimizing the model, we next make a more detailed comparison of model and measurement. 
Fig.~\ref{Fig:SMComp} shows various system matrix rows of the optimized N\'{e}el model B3 in comparison with the equilibrium model and the calibration for various mixing factors and the $x$ receive channel. One can see that the optimized N\'{e}el model much better describes the measurement than the equilibrium model. To support this observation, the figure also shows difference images. Just for $m_x=0$ and $m_x=1,2$ the pattern still looks slightly different and the error is similar for both models. To quantify the differences we calculate the NRMSE for $m_x=0,\dots,9$ and $m_y=0,\dots,9$ which are shown in Fig.~\ref{fig:error}. The data confirms that the N\'{e}el model has lower error than the equilibrium model for most of the mixing factors. In particular for $m_x > m_y$ the error is much lower. For $m_x \ll m_y$, however, the error increases and there is no clear advantage of the N\'{e}el model.

Reconstruction results for the phantom data are shown in Fig.~\ref{fig:reco}. 
One can see that the N\'{e}el model achieves a similar image quality as the calibrated system matrix. 
As expected the results are closer to each other if the threshold $\Theta$ is decreased. 
For large thresholds, the N\'{e}el model B3 results in artifacts at the boundaries of the FoV. 
The equilibrium model A shows a much worse image quality, which underlines, why it is usually not used in practice.

\subsubsection*{I. Discussion} %\paragraph{Discussion}
The results show that the physics of a multi-dimensional MPI experiment can be modeled by solving the Fokker-Planck equation considering N\'{e}el rotation with a spatially inhomogeneous anisotropy. 
%The mean NRMSE of the model on the relevant subset of matrix rows is about 11\% compared to the 15\% for the equilibrium model. 
%Reconstructed data shows that the model shows only slight artifacts compared to a measurement-based reconstruction.
Since the simulation of the N\'{e}el model is computational expensive we did not fully optimize the set of unknown parameters yet. 
One can expect that the accuracy of the model increases even further when considering a particle size distribution and optimizing the probability density function of the particles. 
Instead of the ad-hoc choice of a linear increasing gradient anisotropy it seems also promising to optimize the shape of the anisotropy field. 
The next step for a more precise model could be to consider a coupled  Brownian and N\'{e}el rotation model as for example has been derived in \cite{Weizenecker2018}.

The strengths of using a model compared to using a calibration are manifold. 
We see the prime advantage that it allows to decouple the magnetic field sequence from the particle behavior. 
This enables to use the same model and evaluate it under different field conditions (different excitation-field amplitudes and gradient strength). 
The need for such an approach is even more pressing for multi-patch sequences where a multitude of similar but slightly perturbed system matrices need to be acquired since the fields show imperfections in space \cite{szwargulski2018efficient}.

%In such a scenario one can recalculate the system matrices for different field imperfections without the need for a new calibration measurement.

Another application is multi-contrast MPI \cite{Moeddel_2018}. 
Here, one requires system matrices for different states of the particles (e.g. different viscosity or temperature). 
Using an accurate model, it is possible to determine the required system matrices just by modifying the parameters in the model. 
Furthermore, the problem of multi-contrast MPI can be formulated in terms of a joint parameter identification problem.
%To check the validity of the parameters one might use some selected calibration scans, e.g. at the lowest and highest temperature being images but then one can model the system matrices in between by varying the respective model parameters.

% {\color{blue} Wir sollten diese Studie hier noch einmal diskutieren, da dort gesagt wird, dass fuer Perimag Brown dominiert. Hier sollten wir darauf hinweisen, dass das Problem scheinbar komplexerer Natur ist und das Verhalten womoeglich mit der angelegten Magnetfeldstruktur variiert. Ggf. doch noch den Vergleich mit den solid Messungen heranziehen, die wir gemacht haben, da man dort definitiv die qualitativen Unterschiede sieht (Koennen wir da ggf. einmal auf die Amplituden schauen, bzw. einfch die Colorbar im Vergleich zwischen solid und fluid einmal anschauen? Muss nicht in den Artikel, wuerde mich hier einfach einmal interessiern.) \cite{eberbeck2013multicore} }

%\noindent \textbf{Acknowledgements} 
%T. Kluth acknowledges funding by the Deutsche Forschungsgemeinschaft
%(DFG, German Research Foundation) - project number 281474342/GRK2224/1 ``Pi$^3$ : Parameter Identification
%- Analysis, Algorithms, Applications'' and support by the project ``MPI$^2$'' (BMBF, project no.
%05M16LBA).
%\bibliographystyle{abbrv}
\bibliographystyle{abbrv}
\bibliography{ref}

%\appendix

%\paragraph{Discretization of Fokker-Planck Equation}
%\label{sec:discretizationFP}

%\begin{figure}
%\includegraphics[width=\columnwidth]{pics/20190301_143808.jpg}
%\caption{Vorschlag Phantome}
%\label{fig:Phantome}
%\end{figure}

\end{document}